\documentclass[aps,prl,showpacs,floatfix,twocolumn]{revtex4-1}
\usepackage{graphics,epsfig,graphicx}
\usepackage{amsbsy}
\usepackage{amsmath}
\usepackage{bm}
\usepackage{amsfonts}
\usepackage{cancel}
\usepackage{multirow}
\usepackage{color}

\begin{document}

\title{Universal tetramer limit-cycle at the unitarity limit}

\author{Tobias Frederico}\affiliation{  Instituto Tecnológico de Aeronáutica,
12228-900  S\~ao Jos\'e dos Campos, Brazil}\affiliation{RIKEN, Nishina Center,
Wako, Saitama 351-0198, Japan} 
\author{Mario Gattobigio}\affiliation{  Universit\'e C\^ote d'Azur, CNRS,
Institut  de  Physique  de  Nice,  1361 route des Lucioles, 06560 Valbonne,
France }

\begin{abstract}
  We demonstrate that a four-boson limit-cycle independent of the Efimov one
  appears in Hamiltonian systems at the unitary limit. The model interaction
  contains two-, three- and four-body short-range potentials, which disentangle
  the interwoven three- and four-boson cycles, for the universal trimer and
  tetramer energy levels, respectively. The limit-cycle associated with the
  correlation between the energies of two successive universal tetramer levels
  for fixed weakly bound trimer is found to be largely model independent. This
  is a universal manifestation of an independent four-boson scale associated
  with a cycle beyond the Efimov one.
\end{abstract}
\maketitle

{\it Introduction.\----}~%
It is fascinating how  bosonic quantum systems  behave in the limit of zero
range forces, also known as the scaling limit. Thomas in
1935~\cite{Thomas:1935zz} gave the first hint of its nontrivial properties
showing the  collapse of the three-boson system, meaning the Hamiltonian
spectrum is not bounded from below. Later, Skorniakov and Ter-Matirosyan (STM)
in 1956~\cite{skorniakov1957three}  formulated the three-body integral equations with the
zero-range interaction to be solved, which inspired Faddeev to formulate his famous 
coupled set of three-body integral equations~\cite{faddeev1960scattering}
that overcame the  non-uniqueness problem of the Lippman-Schwinger equations.

However, the solution of the STM equations was plagued by the Thomas collapse,
which was tamed by Danilov in 1961~\cite{danilov1961three}, who found the
log-periodic solutions in the ultraviolet (UV) limit of those equations, as a
consequence of its continuous scale invariance. He recognized the need to
introduce a boundary condition to have a unique solution of the scattering by
giving the three-body binding energy as input. Soon after that, Minlos and
Faddeev~\cite{minlos1962comment} showed how the route to the Thomas collapse of
the bound state proceeds to the "fall-to-center"~\cite{landau2013quantum} by
solving  the homogeneous form of the STM equation for three bosons in the UV
limit, finding an infinite discrete spectrum of the equation that extends to
$-\infty$, with levels geometrically separated by the factor
$\exp(2\pi/s_0)\approx 515$. This was the first clear observation of the
continuous scale symmetry breaking to a discrete one.   

Efimov in 1970~\cite{Efimov:1970zz,Efimov:1971zz} discovered  the presence
of those infinite number of  geometrically spaced levels when a three-boson system interacts
resonantly with any short-range potential, which has an infinite
the scattering length. Nowadays this is called the unitarity
limit in the context of Effective Field Theory (EFT)~\cite{Braaten:2004rn} (see
also~\cite{Hammer:2017tjm}).  

The first experimental evidence of Efimov states came from the Innsbruck
experiment, which used an ultracold gas of cesium
atoms~\cite{kraemer2006evidence}  near a Feshbach resonance. This discovery made
Efimov states a reality, and since then, many other experiments have observed
their presence. Efimov states have also been observed in mass-imbalanced atomic
systems (see e.g.~\cite{Pires:2014zza}).

The Efimov cycle manifests, in practice, through correlations between
observables~\cite{AmorimPRA1999}. It is associated with the three-boson limit
cycle found in the context of EFT~\cite{BedaquePRL1999},
as well as, in other works~\cite{Mohr:2005pv}. 
Moreover, the existence of such correlations is not restricted to zero-range 
interactions, but persists in finite-range systems
too~\cite{kievsky:2013_Phys.Rev.A,kievsky:2015_Phys.Rev.A}.

However, a question arises: how new cycles in addition to the Efimov one
manifest themselves for a larger number of bosons in the unitarity limit, or
in other words, for s-wave interactions in the zero-range limit?

The first nontrivial step is the four-boson system. It was proposed in
Ref.~\cite{Hadizadeh:2011qj}, that a new limit-cycle beyond the Efimov one can
appear in the four-boson system at the unitarity limit. Such a result, expressed by
a correlation between the energies of consecutive tetramers for a fixed trimer
energy, was obtained by solving a regularized set of Faddeev-Yakubovsky (FY)
equations in the limit of a zero-range interaction. The new cycles were revealed
when a four-boson scale was forced to move independently of the three-body one.

In Ref.~\cite{Frederico:2019bnm} it was shown, by an approximate analytic
solution of the FY equations, how  the continuous scale symmetry is broken to a
discrete one, which is associated with a log-periodicity different from the
Efimov one. Such a qualitative view was confirmed within a Born-Oppenheimer
approximation of the heavy-heavy-light-light system, where it was shown
explicitly that the Efimov periodicity of the heavy-heavy-light system is
distinct from the four-body case~\cite{DePaula:2019ryz}, indicating that
new-limit-cycles for more than  three particles are different from the Efimov
cycle at the unitary limit.

While the correlation between consecutive tetramers was supported by the just
mentioned calculations, it remains an open question whether or not a four-boson
Hamiltonian system, with short-range forces at the unitarity limit, could
exhibit a typical four-boson limit cycle independent of the three-boson one.
The four-boson cycle was not seen before in Hamiltonian systems when using two
and three-body forces, which exhibited for each Efimov state two
tetramers~\cite{Platter:2004he,StecherNat2009,GattobigioPRA2011,Gattobigio:2012yky,GattobigioPRA2014,Kievsky:2014yua,KievskyPRA2014,RodriguezPRA2016}.
However, an indication that this independent limit-cycle could exist in
Hamiltonian systems comes from the accurate calculations provided in
Ref.~\cite{Deltuva:2012ig}, which also suggests that the property of the Efimov
cycle being interwoven with the four-boson one is verified as pointed out
in~\cite{Hadizadeh:2011qj} and confirmed in Ref.~\cite{DePaula:2019ryz} for
heavy-heavy-light-light bosonic systems and beyond, within the Born-Oppenheimer
treatment.

A crucial property associated with the independent three and four-boson
limit-cycles is the necessity of the introduction of a four-boson scale
unrelated to the three-boson one. Such possibility introduced
in~\cite{Yamashita_2006} confronted previous findings within
EFT~\cite{Platter:2004he} and further explored in Ref.~\cite{StecherNat2009},
which by now was understood that at next-to-leading order (NLO) a four-boson
scale has to be introduced in the EFT approach of the universal
system~\cite{Bazak:2018qnu}.  The appearance of a four-body scale was also
confirmed in calculations up to five bosons  with van der Waals
interactions~\cite{Stipanovic2022}. Therefore, it is appealing to study the
eigenvalues of  a four-boson Hamiltonian for short-range interactions at the
unitarity limit, with two-, three- and four-body potentials to disentangle the
three and four-boson cycles~\cite{horinouchi:2015_Phys.Rev.Lett.} by
manipulating  the three and four-boson short-range scales in an independent
way, to follow the path of the recognized two universal tetramer levels
attached to an Efimov state found in Hamiltonian
models~\cite{Platter:2004he,StecherNat2009,Gattobigio:2012yky,GattobigioPRA2014}.
 
 Evidence of
induced multi-boson interactions~\cite{Yamashita_2006,Yamashita:2020jvu} may be
already found in cold atomic gases, where the position of  three-atom resonances
for narrow Feshbach resonances~\cite{ChinNatPhys2017} and also for intermediate
ones~\cite{Chapurin_2019,CornellPRL2020} deviate significantly from the predictions based on
the van der Waals universality (see e.g.~\cite{Naidon:2016dpf,Greene:2017cik}).

In this work, we demonstrate that the limit-cycle for the correlation
between successive universal tetramers energies is present in a Hamiltonian system for a fixed
trimer energy at unitarity. Our results are in agreement, taking into account
the range corrections, with the predictions of Ref.\cite{Hadizadeh:2011qj},
which relies on the solution of the regularized four-boson FY equations. The
Hamiltonian used in our investigation incorporates two-, three- and four-body
short-range potentials. The two-body potential is employed to adjust the
two-boson energy at the unitarity point, while the three- and four-body
potential is utilized to disentangle the interwoven three-boson cycle and the elusive four-boson one, whose existence
has been a subject of debate in the
field~\cite{Platter:2004he,Yamashita_2006,Hadizadeh:2011qj,StecherNat2009,Bazak:2018qnu}.

{\it The Hamiltonian Model.\----}~
We  solve variationally the Schr\"odinger equation for a system of three and four
bosons finding the ground state and several excited states. We develop the
states using a set of correlated Gaussian functions, which were previously
optimized using the Stochastic Variational
Method (SVM)~\cite{varga:1995_Phys.Rev.C,suzuki:1998_}.  SVM is essential in adapting
the basis functions to different scales, which is a crucial factor in the
calculation of multiple excited states.

As we want to study the behavior of systems interacting via a short-range
interaction at the unitary limit, we choose the Gaussian potential as a
representative of such interactions. It has been extensively shown that close to
the unitarity limit the Gaussian potentials gives a universal representation of the
class of short-range
potentials~\cite{deltuva:2020_Phys.Rev.C,kievsky:2021_Annu.Rev.Nucl.Part.Sci.}.
The potential we use has a two-, three-, and four-particle Gaussian terms. The
two-body term reads
\begin{equation}
  V_{2b}(r) = V_2 \, e^{-(r/r_2)^2} \,,
  \label{eq:2bGauss}
\end{equation}
and depends on the relative distance $r$ between two particles. The
range of the force is fixed to $r_2=1$, which it is used as unit of
length; in this way the energy unit
is $\hbar^2/mr_2^2$.
The two-body strength is set to $V_2 = -2.684005~\hbar^2/mr_2^2$ to tune
the two-body system as close to the unitarity limit as possible.

The three-body term is
\begin{equation}
  V_{3b}(r) = V_3 \, e^{-(\rho_3/r_3)^2}\,,
  \label{eq:3bGauss}
\end{equation}
where $\rho_3^2 = r_{12}^2 + r_{13}^2 + r_{23}^2$ is proportional to the
three-body hyper-radius, and $r_3$ is the potential range in units of $r_2$. This
three-body force can be used to change the value of the three-body ground state
energy.

To accomplish our goal, we introduce a four-body potential
\begin{equation}
  V_{4b}(r) = V_4 \, e^{-(\rho_4/r_4)^2}\,,
  \label{eq:4bGauss}
\end{equation}
where $\rho_4^2 = \sum_{i<j=1}^4 r_{ij}^2$ is proportional to the four-body
hyper-radius, and $r_4$ is the potential range in units of $r_2$. We use this
four-body force to change the energy of the four-body states below the
three-body threshold. Moreover, it is also used to change the number of
four-body states below the threshold. With the ensemble of these
forces, we can address individuality the two-, three- and four-body
energy levels.

\begin{figure}[t]
  \includegraphics[width=\linewidth]{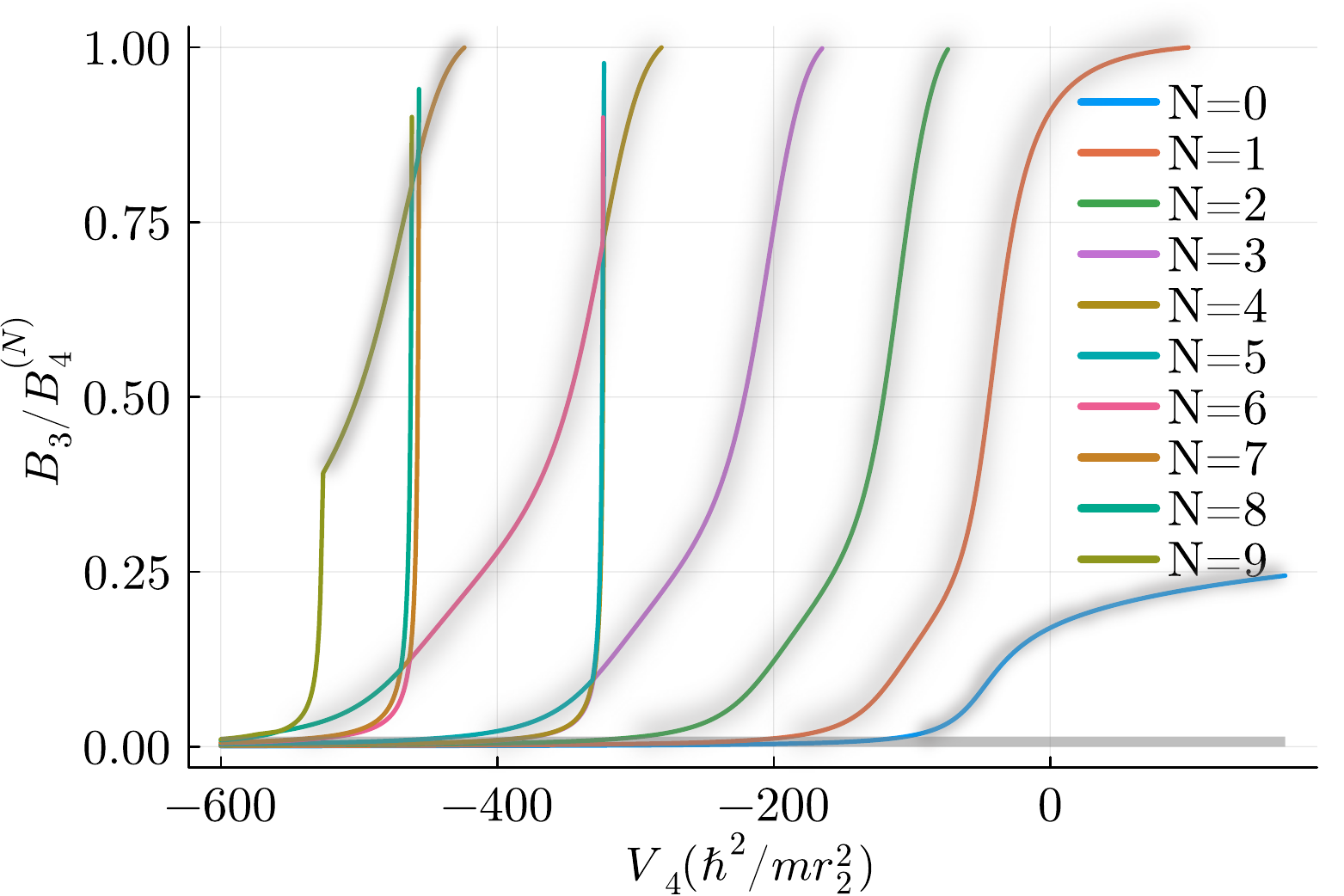}
  \includegraphics[width=\linewidth]{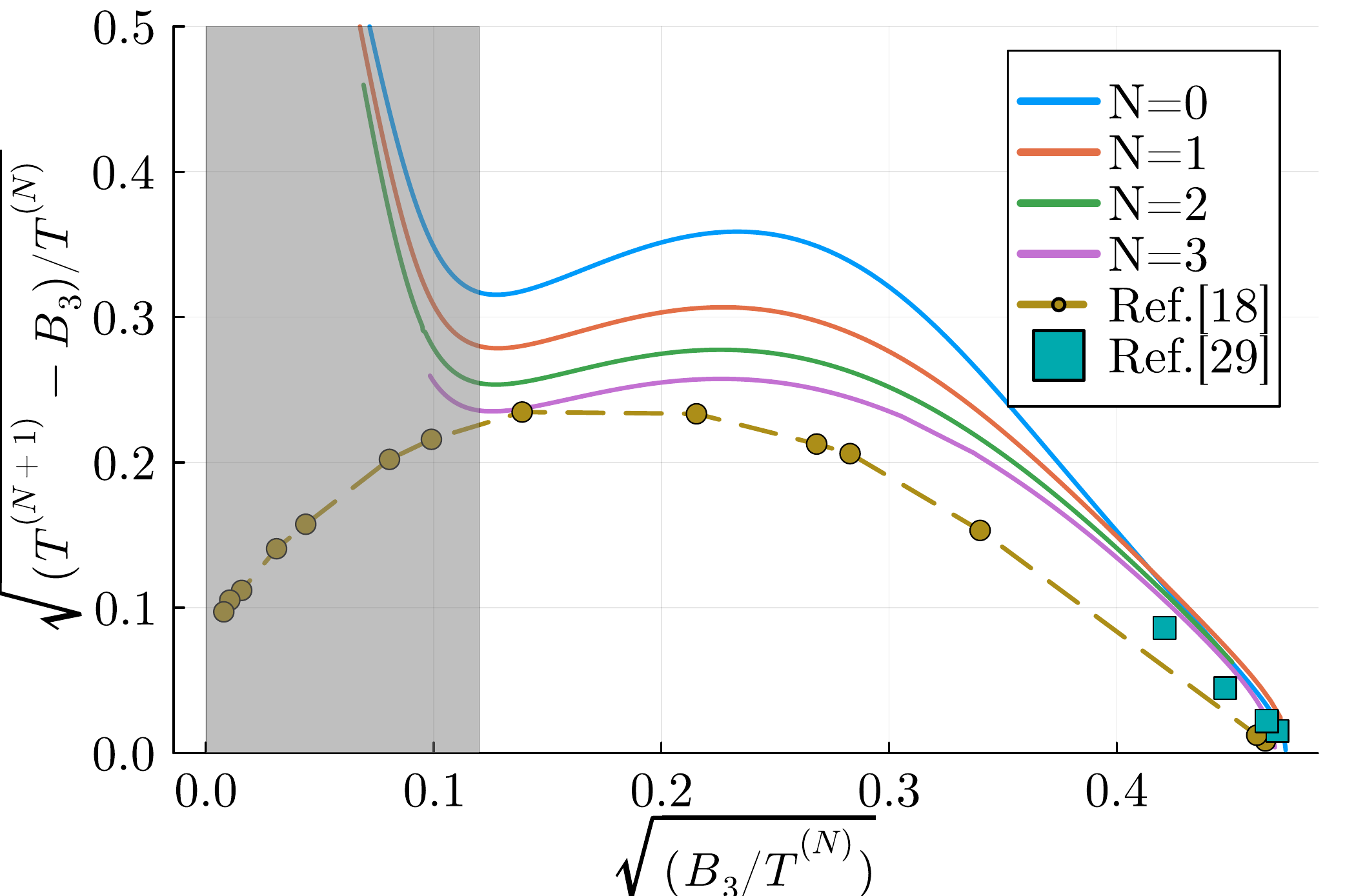}
 
  \caption{Top panel: four-bosons energy levels as a function of the
    strength of the four-body force $V_4$. The three-boson
    bound-state energy is $B_3 =0.238451~\hbar^2/mr_2^2$, while the two-body
    system is at the unitarity limit. The four-body potential range is $r_4=r_2$.
    Bottom panel: the cycle for  four bosons obtained by fixing the two-body system
    at the unitarity point, $B_2=0$, and varying the four-body strength $V_4$.  In
    the same graph we have the cycle involving the ground and the first-excited
    stated, $N=0$, and the cycle involving the
    first- and the second-excited state, $N=1$, and so on up to $N=3$.}
  \label{fig:ourCycle}
\end{figure}

{\it Results.\----}~%
Our Hamiltonian model with Gaussian potentials provides evidence of the
existence of cycles for the universal four-boson states independent of the
Efimov cycles, which are presented in what follows.

Our calculations are shown in Fig.~\ref{fig:ourCycle}, where we set the
two-body strength to $V_2 = -2.684005~\hbar^2/mr_2^2$ to ensure $B_2=0$, and set
the three-body force to zero ($V_3=0$). Using these parameter values, the binding energy
of the three-boson bound-state is $B_3=0.238451~\hbar^2/mr_2^2$.

After fixing the two- and three-body sectors, we use the four-body force to
manipulate the four-body spectrum, identifying the universal levels and examining
correlations among these states.  Specifically, we vary $V_4$ while keeping the
range of the four-body force fixed to $r_4=r_2$. The resulting spectrum is
depicted in the top panel of Fig.~\ref{fig:ourCycle}, where we plot the ratio
$B_3/B_4^{(N)}$ against $V_4$.

As we increase the strength of $V_4$, more states gradually appear in the
spectrum, emerging from the three-body threshold. Interestingly, we observe two
types of states: the first ($N=1$) and second excited states ($N=2$) smoothly
move to deeper values after emerging from the threshold. However, there are
states like the third excited state ($N=3$) that also emerge smoothly from the
threshold, but at a certain point, such as at $V_4\approx -323~\hbar^2/mr_2^2$
in this case, they exhibit a strong avoided crossing with other states, for
instance, with $N=4$, and $N=5$. The state that emerges from the threshold as
$N=5$ evolves in a much narrower range of the four-body interaction, first
interacting with the $N=4$ and $N=6$ states and then with $N=3$ and $N=4$.
During this interaction, there is a role exchange, and we can still trace the
smooth trajectory of the $N=3$ state, but now as $N=5$. This is just one example
of the avoided-crossing structure of the spectrum, which becomes more evident as
we increase the strength of the four-body force.

The avoided-crossing structure arises due to the interplay between the
long-range effective (hyper-radial) potential, similar to the three-body
system~\cite{Efimov:1970zz}, and the short-range nature of the four-body force
used to reveal the universal states and their underlying correlations. This
competition between the short- and long-range interactions results in the
avoided-crossing structure of the spectrum. This phenomenon has been
extensively studied in the case of exotic
atoms~\cite{combescure:2007_Int.J.Mod.Phys.B}.

In terms of studying correlations, we only consider the energy levels resulting
from the long-range interaction and therefore having universal scaling
properties.  We use their energy, denoted as $T^{(N)}$, to construct the
correlation, as shown in the bottom panel of Fig.~\ref{fig:ourCycle}. To
clarify, in the case of Fig.~\ref{fig:ourCycle}, we have $T^{(0)} = B_4^{(0)}$,
$T^{(1)} = B_4^{(1)}$, and $T^{(2)} = B_4^{(2)}$.  However, for $T^{(3)}$, it is
equal to $B_4^{(3)}$ only up to $V_4\approx -323\hbar^2/mr_2^2$, after which it
becomes equal to $B_4^{(5)}$. In the top panel of Fig.~\ref{fig:ourCycle}, these
states have been highlighted.

The correlation function between the energies of two consecutive universal
tetramers is plotted in the bottom panel of Fig.\ref{fig:ourCycle} using the
$T^{(N)}$ levels. This function is constructed as suggested in
Ref.~\cite{Hadizadeh:2011qj} by plotting $\sqrt{(T^{(N+1)}-B_3)/T^{(N)}}$ as a
function of $\sqrt{B_3/T^{(N)}}$. The resulting plot demonstrates the
limit-cycle consistently. We display four such cycles, and we observe that they
exhibit a convergence pattern as a function of $N$. While the difference between
the cycles is more pronounced for $\sqrt{B_3/T^{(N)}}\approx 0.25$, they
collapse to the same curve for $\sqrt{B_3/T^{(N)}}\approx 0.45$, which
corresponds to the point where a new tetramer emerges from the three-body
threshold, and the state with energy $T^{(N)}$ is sufficiently shallow.

For the sake of comparison, in the same plot, we report the zero-range
calculation of Hadizadeh et al~\cite{Hadizadeh:2011qj}, and also the results
from Deltuva~\cite{Deltuva:2012ig} obtained with separable potentials for
tetramers associated with different trimers at the unitarity limit. The general
trend of the limit-cycle obtained with the Gaussian potentials reproduces both
the zero-range model and also the separable calculations, which are placed
close to the point where one of the tetramers hits the trimer threshold.  We
believe that the difference between the zero-range cycle and the present
results is possibly due to  range corrections.

Moreover, we  observe another  noticeable discrepancy between the zero-range
and finite-range calculations when examining values lower than
$\sqrt{B_3/T^{(N)}}\approx 0.12$. Instead of continuing to decrease, the
correlation cycles begin to grow in this region, which is covered by the shadow
box. As a result, the $T^{(N)}$ state transitions from the universal window
into a very tight four-body state. The same gray region is depicted in the
Fig.~\ref{fig:ourCycle} top panel.

\begin{figure}[t]

\includegraphics[width=\linewidth]{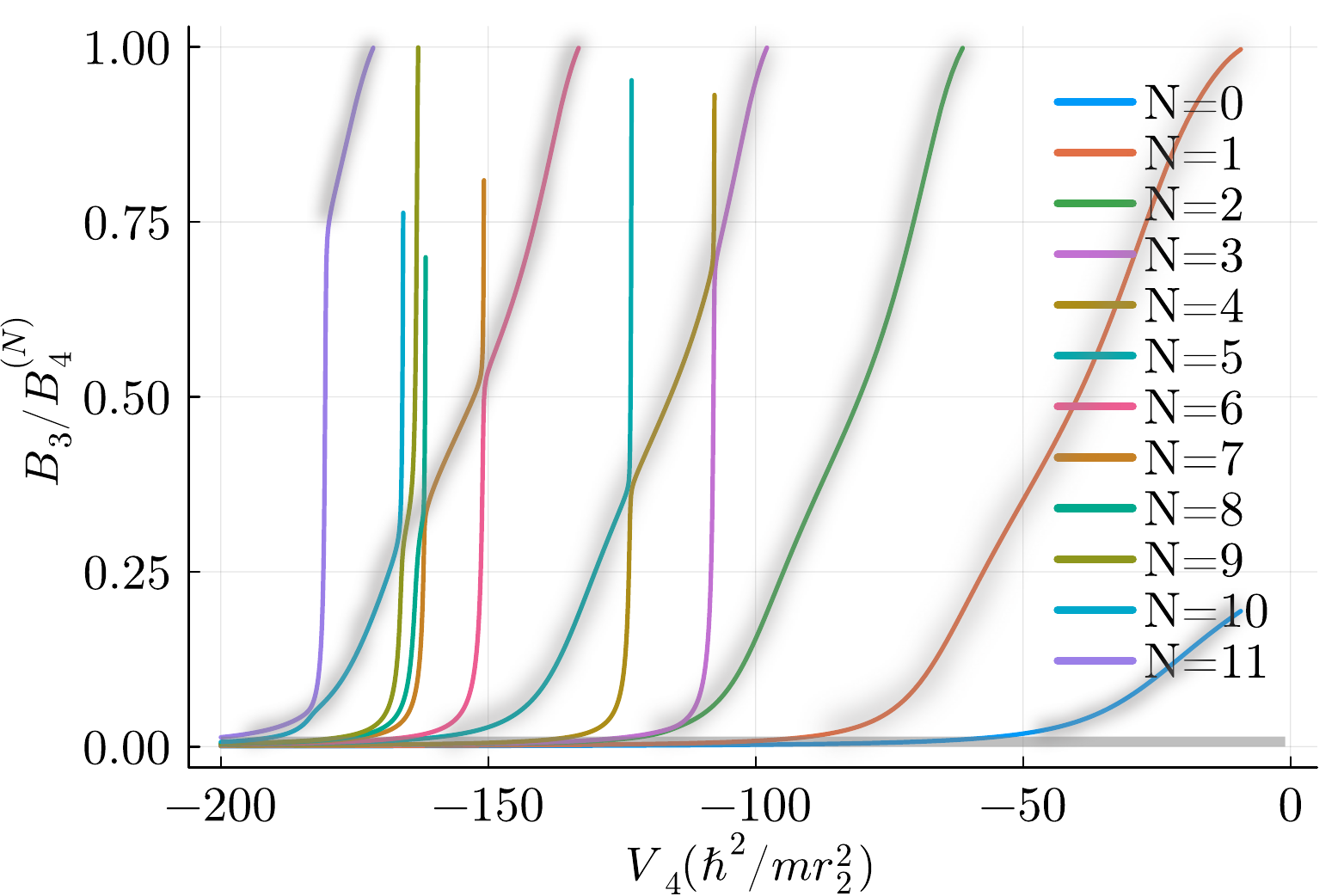}
  \includegraphics[width=\linewidth]{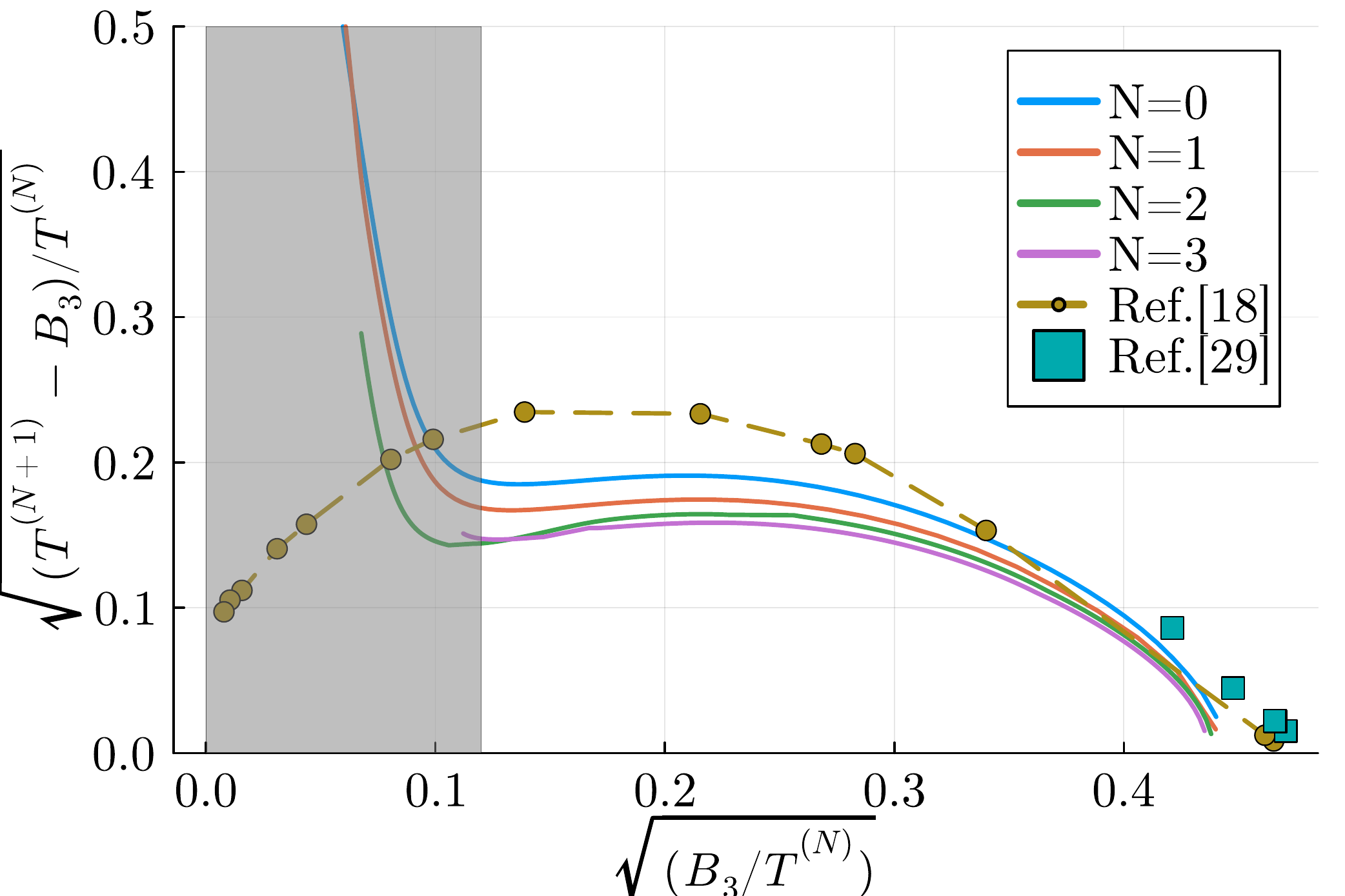}
  
  \caption{Top panel: Evolution of the first eleven energy levels  as a
    function of the strength $V_4$, obtained fixing the two-body system at the
    unitary point, $B_2=0$. In this case, a three-body force has been
    introduced so that the three-body bound state energy is $B_3=
    0.04001~\hbar^2/mr_2^2$ and the four-body potential range is $r_4=2r_2$.
    We observe that starting from the third-excited four-boson state, $N=3$,
    there is the appearance of the avoided crossing, and the evolution of the
    universal levels is evidenced by the shadow line. Bottom panel: The cycle
    for the  four-bosons successive universal levels.  We have four cycles that
    have been obtained by following the universal states through the avoided
    crossing.}
  \label{fig:threeCycles}
\end{figure}

The finite-range form of the universal limit-cycle in the four-boson system
depends also on the value of the three-boson energy, which can be changed using
a non-vanishing three-body force. To explore this dependence, we set the
three-body strength to $V_3=30~\hbar^2/mr_2^2$ and $r_3=r_2$, which shifts the
trimer to a weakly bound state with energy $B_3=0.04001~\hbar^2/mr_2^2$. Using a
four-body range of $r_4=2r_2$, we perform new calculations, and the results are
reported in Fig.~\ref{fig:threeCycles}. On the top panel, we show the energy
levels, which display the same avoided crossing pattern as in the previous case.
We also identify the universal tetramers $T^{(N)}$, highlighted in the same
panel, which show avoided crossing starting from the third-excited level $N=3$.
The correlation between these states is shown in the bottom panel of
Fig.~\ref{fig:threeCycles}, along with results from previous studies,
Refs.~\cite{Hadizadeh:2011qj} and \cite{Deltuva:2012ig}, for comparison.

Although the correlations are similar to the case with the tighter trimer, there
are differences, such as a lower maximum value, which can be attributed to the
finite-range nature of the interaction.  However, the trend of the limit-cycle
obtained with the Gaussian potentials for the universal tetramer levels is
closer to both the zero-range model and the separable potential calculations,
which are located near the point where one of the tetramers hits the trimer
threshold. The better agreement with the zero-range results is evident in the
interval where $\sqrt{(B_3/T^{(N)})}\gtrsim 0.3$, although range corrections are
still present. The gray region in both panels of Fig.~\ref{fig:threeCycles}
represents the strongly bound tetramers, which are outside the "window" where
the universal states are. One way to modify the finite-range effects is by
changing the range of the four-body force $r_4$. We checked that while this
change the structure of the energy levels, the correlations remain the same.

Despite these finite-range details, we have successfully proven the existence of
a universal tetramer limit-cycle independent of the trimer; and we leave for
future research more refined Hamiltonian calculations that can control finite
range effects.

{\it Summary.\----}~% 
We have demonstrated the unexpected presence of a universal limit-cycle in
Hamiltonian systems, which arises from the breaking of continuous scale symmetry
to a residual discrete scale invariance in four-boson systems. This phenomenon
is independent of the one appearing in the three-boson system. Our Hamiltonian
system consists of two-, three-, and four-body short-range interactions, tuned
to the unitarity limit. We have identified a series of universal tetramer states
that maintain their model-independent properties, including the correlation
between the energies of successive levels that converge toward the limit-cycle.
Our results suggest that the use of Gaussian interactions does not restrict the
exploration of these new cycles, as different parametrizations demonstrate the
persistence of the universal levels and the associated energy correlation.

Our work opens the surprising perspective to search for new interwoven
limit-cycles in the $N\geq 5$ boson systems, by exploring the rich pattern
arising when moving the universal levels through the control of many-body
potentials.
How this can be done in practice, for instance in cold-atom experiments, is an open
question.  Recently, the possibility of many-body forces manifesting
themselves near relatively narrow Feshbach resonances has been suggested, which
is attributed to the coupling between the open and closed channels when the
atom-atom interaction is magnetically tuned in cold traps. However, further
research is needed to explore this possibility, as highlighted in recent studies
such as~\cite{KokkelmansPRA2021,Yamashita:2020jvu}.

  {\it Acknowledgements.\----}~%
  We are thankful to Alejandro Kievsky for the suggestion of the use of
  four-body potentials in this problem.  TF  thanks the warm hospitality of
  Emiko Hiyama at the Tohoku University and at the Nishina Center in Riken.  TF
  thanks for the support by Conselho Nacional de Desenvolvimento Cient\'ifico e
  Tecnol\'ogico (CNPq) under the grants 308486/2022-7  and INCT-FNA project
  464898/2014-5, and by Funda\c c\~ao de Amparo \`a Pesquisa do Estado de S\~ao
  Paulo (FAPESP) Thematic grants 2017/05660-0 and 2019/07767-1, and
  CAPES/COFECUB (Coordena\c c\~ao de Aperfeiçoamento de Pessoal de N\'ivel
  Superior) grant 88887.370819/2019-0.  
 This research was boosted during the  program "Living Near Unitarity" at the
 Kavli Institute for Theoretical Physics (KITP),  University of Santa Barbara
 (California) and supported in part by the National Science Foundation under
 Grant No. NSF PHY-1748958.


\begin{thebibliography}{45}%
\makeatletter
\providecommand \@ifxundefined [1]{%
 \@ifx{#1\undefined}
}%
\providecommand \@ifnum [1]{%
 \ifnum #1\expandafter \@firstoftwo
 \else \expandafter \@secondoftwo
 \fi
}%
\providecommand \@ifx [1]{%
 \ifx #1\expandafter \@firstoftwo
 \else \expandafter \@secondoftwo
 \fi
}%
\providecommand \natexlab [1]{#1}%
\providecommand \enquote  [1]{``#1''}%
\providecommand \bibnamefont  [1]{#1}%
\providecommand \bibfnamefont [1]{#1}%
\providecommand \citenamefont [1]{#1}%
\providecommand \href@noop [0]{\@secondoftwo}%
\providecommand \href [0]{\begingroup \@sanitize@url \@href}%
\providecommand \@href[1]{\@@startlink{#1}\@@href}%
\providecommand \@@href[1]{\endgroup#1\@@endlink}%
\providecommand \@sanitize@url [0]{\catcode `\\12\catcode `\$12\catcode
  `\&12\catcode `\#12\catcode `\^12\catcode `\_12\catcode `\%12\relax}%
\providecommand \@@startlink[1]{}%
\providecommand \@@endlink[0]{}%
\providecommand \url  [0]{\begingroup\@sanitize@url \@url }%
\providecommand \@url [1]{\endgroup\@href {#1}{\urlprefix }}%
\providecommand \urlprefix  [0]{URL }%
\providecommand \Eprint [0]{\href }%
\providecommand \doibase [0]{http://dx.doi.org/}%
\providecommand \selectlanguage [0]{\@gobble}%
\providecommand \bibinfo  [0]{\@secondoftwo}%
\providecommand \bibfield  [0]{\@secondoftwo}%
\providecommand \translation [1]{[#1]}%
\providecommand \BibitemOpen [0]{}%
\providecommand \bibitemStop [0]{}%
\providecommand \bibitemNoStop [0]{.\EOS\space}%
\providecommand \EOS [0]{\spacefactor3000\relax}%
\providecommand \BibitemShut  [1]{\csname bibitem#1\endcsname}%
\let\auto@bib@innerbib\@empty
%</preamble>
\bibitem [{\citenamefont {Thomas}(1935)}]{Thomas:1935zz}%
  \BibitemOpen
  \bibfield  {author} {\bibinfo {author} {\bibfnamefont {L.~H.}\ \bibnamefont
  {Thomas}},\ }\href {\doibase 10.1103/PhysRev.47.903} {\bibfield  {journal}
  {\bibinfo  {journal} {Phys. Rev.}\ }\textbf {\bibinfo {volume} {47}},\
  \bibinfo {pages} {903} (\bibinfo {year} {1935})}\BibitemShut {NoStop}%
\bibitem [{\citenamefont {Skorniakov}\ and\ \citenamefont
  {Ter-Martirosian}(1957)}]{skorniakov1957three}%
  \BibitemOpen
  \bibfield  {author} {\bibinfo {author} {\bibfnamefont {G.}~\bibnamefont
  {Skorniakov}}\ and\ \bibinfo {author} {\bibfnamefont {K.}~\bibnamefont
  {Ter-Martirosian}},\ }\href@noop {} {\bibfield  {journal} {\bibinfo
  {journal} {Soviet Phys. JETP}\ }\textbf {\bibinfo {volume} {4}} (\bibinfo
  {year} {1957})}\BibitemShut {NoStop}%
\bibitem [{\citenamefont {Faddeev}(1960)}]{faddeev1960scattering}%
  \BibitemOpen
  \bibfield  {author} {\bibinfo {author} {\bibfnamefont {L.}~\bibnamefont
  {Faddeev}},\ }\href@noop {} {\bibfield  {journal} {\bibinfo  {journal} {Zhur.
  Eksptl'. i Teoret. Fiz.}\ }\textbf {\bibinfo {volume} {39}} (\bibinfo {year}
  {1960})}\BibitemShut {NoStop}%
\bibitem [{\citenamefont {Danilov}(1961)}]{danilov1961three}%
  \BibitemOpen
  \bibfield  {author} {\bibinfo {author} {\bibfnamefont {G.}~\bibnamefont
  {Danilov}},\ }\href@noop {} {\bibfield  {journal} {\bibinfo  {journal} {Sov.
  Phys. JETP}\ }\textbf {\bibinfo {volume} {13}},\ \bibinfo {pages} {3}
  (\bibinfo {year} {1961})}\BibitemShut {NoStop}%
\bibitem [{\citenamefont {Minlos}\ and\ \citenamefont
  {Faddeev}(1962)}]{minlos1962comment}%
  \BibitemOpen
  \bibfield  {author} {\bibinfo {author} {\bibfnamefont {R.}~\bibnamefont
  {Minlos}}\ and\ \bibinfo {author} {\bibfnamefont {L.}~\bibnamefont
  {Faddeev}},\ }\href@noop {} {\bibfield  {journal} {\bibinfo  {journal} {Sov.
  Phys. JETP}\ }\textbf {\bibinfo {volume} {14}},\ \bibinfo {pages} {1315}
  (\bibinfo {year} {1962})}\BibitemShut {NoStop}%
\bibitem [{\citenamefont {Landau}\ and\ \citenamefont
  {Lifshitz}(2013)}]{landau2013quantum}%
  \BibitemOpen
  \bibfield  {author} {\bibinfo {author} {\bibfnamefont {L.~D.}\ \bibnamefont
  {Landau}}\ and\ \bibinfo {author} {\bibfnamefont {E.~M.}\ \bibnamefont
  {Lifshitz}},\ }\href@noop {} {\emph {\bibinfo {title} {Quantum mechanics:
  non-relativistic theory}}},\ Vol.~\bibinfo {volume} {3}\ (\bibinfo
  {publisher} {Elsevier},\ \bibinfo {year} {2013})\BibitemShut {NoStop}%
\bibitem [{\citenamefont {Efimov}(1970)}]{Efimov:1970zz}%
  \BibitemOpen
  \bibfield  {author} {\bibinfo {author} {\bibfnamefont {V.}~\bibnamefont
  {Efimov}},\ }\href {\doibase 10.1016/0370-2693(70)90349-7} {\bibfield
  {journal} {\bibinfo  {journal} {Phys. Lett. B}\ }\textbf {\bibinfo {volume}
  {33}},\ \bibinfo {pages} {563} (\bibinfo {year} {1970})}\BibitemShut
  {NoStop}%
\bibitem [{\citenamefont {Efimov}(1971)}]{Efimov:1971zz}%
  \BibitemOpen
  \bibfield  {author} {\bibinfo {author} {\bibfnamefont {V.~N.}\ \bibnamefont
  {Efimov}},\ }\href@noop {} {\bibfield  {journal} {\bibinfo  {journal} {Sov.
  J. Nucl. Phys.}\ }\textbf {\bibinfo {volume} {12}},\ \bibinfo {pages} {589}
  (\bibinfo {year} {1971})}\BibitemShut {NoStop}%
\bibitem [{\citenamefont {Braaten}\ and\ \citenamefont
  {Hammer}(2006)}]{Braaten:2004rn}%
  \BibitemOpen
  \bibfield  {author} {\bibinfo {author} {\bibfnamefont {E.}~\bibnamefont
  {Braaten}}\ and\ \bibinfo {author} {\bibfnamefont {H.~W.}\ \bibnamefont
  {Hammer}},\ }\href {\doibase 10.1016/j.physrep.2006.03.001} {\bibfield
  {journal} {\bibinfo  {journal} {Phys. Rept.}\ }\textbf {\bibinfo {volume}
  {428}},\ \bibinfo {pages} {259} (\bibinfo {year} {2006})},\ \Eprint
  {http://arxiv.org/abs/cond-mat/0410417} {arXiv:cond-mat/0410417} \BibitemShut
  {NoStop}%
\bibitem [{\citenamefont {Hammer}\ \emph {et~al.}(2017)\citenamefont {Hammer},
  \citenamefont {Ji},\ and\ \citenamefont {Phillips}}]{Hammer:2017tjm}%
  \BibitemOpen
  \bibfield  {author} {\bibinfo {author} {\bibfnamefont {H.~W.}\ \bibnamefont
  {Hammer}}, \bibinfo {author} {\bibfnamefont {C.}~\bibnamefont {Ji}}, \ and\
  \bibinfo {author} {\bibfnamefont {D.~R.}\ \bibnamefont {Phillips}},\ }\href
  {\doibase 10.1088/1361-6471/aa83db} {\bibfield  {journal} {\bibinfo
  {journal} {J. Phys. G}\ }\textbf {\bibinfo {volume} {44}},\ \bibinfo {pages}
  {103002} (\bibinfo {year} {2017})},\ \Eprint
  {http://arxiv.org/abs/1702.08605} {arXiv:1702.08605 [nucl-th]} \BibitemShut
  {NoStop}%
\bibitem [{\citenamefont {N{\"a}gerl}\ \emph {et~al.}(2006)\citenamefont
  {N{\"a}gerl} \emph {et~al.}}]{kraemer2006evidence}%
  \BibitemOpen
  \bibfield  {author} {\bibinfo {author} {\bibfnamefont {H.-C.}\ \bibnamefont
  {N{\"a}gerl}} \emph {et~al.},\ }\href@noop {} {\bibfield  {journal} {\bibinfo
   {journal} {Nature}\ }\textbf {\bibinfo {volume} {440}},\ \bibinfo {pages}
  {315} (\bibinfo {year} {2006})}\BibitemShut {NoStop}%
\bibitem [{\citenamefont {Pires}\ \emph {et~al.}(2014)\citenamefont {Pires},
  \citenamefont {Ulmanis}, \citenamefont {H\"afner}, \citenamefont {Repp},
  \citenamefont {Arias}, \citenamefont {Kuhnle},\ and\ \citenamefont
  {Weidem\"uller}}]{Pires:2014zza}%
  \BibitemOpen
  \bibfield  {author} {\bibinfo {author} {\bibfnamefont {R.}~\bibnamefont
  {Pires}}, \bibinfo {author} {\bibfnamefont {J.}~\bibnamefont {Ulmanis}},
  \bibinfo {author} {\bibfnamefont {S.}~\bibnamefont {H\"afner}}, \bibinfo
  {author} {\bibfnamefont {M.}~\bibnamefont {Repp}}, \bibinfo {author}
  {\bibfnamefont {A.}~\bibnamefont {Arias}}, \bibinfo {author} {\bibfnamefont
  {E.~D.}\ \bibnamefont {Kuhnle}}, \ and\ \bibinfo {author} {\bibfnamefont
  {M.}~\bibnamefont {Weidem\"uller}},\ }\href {\doibase
  10.1103/PhysRevLett.112.250404} {\bibfield  {journal} {\bibinfo  {journal}
  {Phys. Rev. Lett.}\ }\textbf {\bibinfo {volume} {112}},\ \bibinfo {pages}
  {250404} (\bibinfo {year} {2014})},\ \Eprint {http://arxiv.org/abs/1403.7246}
  {arXiv:1403.7246 [cond-mat.quant-gas]} \BibitemShut {NoStop}%
\bibitem [{\citenamefont {Frederico}\ \emph {et~al.}(1999)\citenamefont
  {Frederico}, \citenamefont {Tomio}, \citenamefont {Delfino},\ and\
  \citenamefont {Amorim}}]{AmorimPRA1999}%
  \BibitemOpen
  \bibfield  {author} {\bibinfo {author} {\bibfnamefont {T.}~\bibnamefont
  {Frederico}}, \bibinfo {author} {\bibfnamefont {L.}~\bibnamefont {Tomio}},
  \bibinfo {author} {\bibfnamefont {A.}~\bibnamefont {Delfino}}, \ and\
  \bibinfo {author} {\bibfnamefont {A.~E.~A.}\ \bibnamefont {Amorim}},\ }\href
  {\doibase 10.1103/PhysRevA.60.R9} {\bibfield  {journal} {\bibinfo  {journal}
  {Phys. Rev. A}\ }\textbf {\bibinfo {volume} {60}},\ \bibinfo {pages} {R9}
  (\bibinfo {year} {1999})}\BibitemShut {NoStop}%
\bibitem [{\citenamefont {Bedaque}\ \emph {et~al.}(1999)\citenamefont
  {Bedaque}, \citenamefont {Hammer},\ and\ \citenamefont {van
  Kolck}}]{BedaquePRL1999}%
  \BibitemOpen
  \bibfield  {author} {\bibinfo {author} {\bibfnamefont {P.~F.}\ \bibnamefont
  {Bedaque}}, \bibinfo {author} {\bibfnamefont {H.-W.}\ \bibnamefont {Hammer}},
  \ and\ \bibinfo {author} {\bibfnamefont {U.}~\bibnamefont {van Kolck}},\
  }\href {\doibase 10.1103/PhysRevLett.82.463} {\bibfield  {journal} {\bibinfo
  {journal} {Phys. Rev. Lett.}\ }\textbf {\bibinfo {volume} {82}},\ \bibinfo
  {pages} {463} (\bibinfo {year} {1999})}\BibitemShut {NoStop}%
\bibitem [{\citenamefont {Mohr}\ \emph {et~al.}(2006)\citenamefont {Mohr},
  \citenamefont {Furnstahl}, \citenamefont {Perry}, \citenamefont {Wilson},\
  and\ \citenamefont {Hammer}}]{Mohr:2005pv}%
  \BibitemOpen
  \bibfield  {author} {\bibinfo {author} {\bibfnamefont {R.~F.}\ \bibnamefont
  {Mohr}}, \bibinfo {author} {\bibfnamefont {R.~J.}\ \bibnamefont {Furnstahl}},
  \bibinfo {author} {\bibfnamefont {R.~J.}\ \bibnamefont {Perry}}, \bibinfo
  {author} {\bibfnamefont {K.~G.}\ \bibnamefont {Wilson}}, \ and\ \bibinfo
  {author} {\bibfnamefont {H.~W.}\ \bibnamefont {Hammer}},\ }\href {\doibase
  10.1016/j.aop.2005.10.002} {\bibfield  {journal} {\bibinfo  {journal} {Annals
  Phys.}\ }\textbf {\bibinfo {volume} {321}},\ \bibinfo {pages} {225} (\bibinfo
  {year} {2006})},\ \Eprint {http://arxiv.org/abs/nucl-th/0509076}
  {arXiv:nucl-th/0509076} \BibitemShut {NoStop}%
\bibitem [{\citenamefont {Kievsky}\ and\ \citenamefont
  {Gattobigio}(2013)}]{kievsky:2013_Phys.Rev.A}%
  \BibitemOpen
  \bibfield  {author} {\bibinfo {author} {\bibfnamefont {A.}~\bibnamefont
  {Kievsky}}\ and\ \bibinfo {author} {\bibfnamefont {M.}~\bibnamefont
  {Gattobigio}},\ }\href {\doibase 10.1103/PhysRevA.87.052719} {\bibfield
  {journal} {\bibinfo  {journal} {Phys. Rev. A}\ }\textbf {\bibinfo {volume}
  {87}},\ \bibinfo {pages} {052719} (\bibinfo {year} {2013})}\BibitemShut
  {NoStop}%
\bibitem [{\citenamefont {Kievsky}\ and\ \citenamefont
  {Gattobigio}(2015)}]{kievsky:2015_Phys.Rev.A}%
  \BibitemOpen
  \bibfield  {author} {\bibinfo {author} {\bibfnamefont {A.}~\bibnamefont
  {Kievsky}}\ and\ \bibinfo {author} {\bibfnamefont {M.}~\bibnamefont
  {Gattobigio}},\ }\href {\doibase 10.1103/PhysRevA.92.062715} {\bibfield
  {journal} {\bibinfo  {journal} {Phys. Rev. A}\ }\textbf {\bibinfo {volume}
  {92}},\ \bibinfo {pages} {062715} (\bibinfo {year} {2015})}\BibitemShut
  {NoStop}%
\bibitem [{\citenamefont {Hadizadeh}\ \emph {et~al.}(2011)\citenamefont
  {Hadizadeh}, \citenamefont {Yamashita}, \citenamefont {Tomio}, \citenamefont
  {Delfino},\ and\ \citenamefont {Frederico}}]{Hadizadeh:2011qj}%
  \BibitemOpen
  \bibfield  {author} {\bibinfo {author} {\bibfnamefont {M.~R.}\ \bibnamefont
  {Hadizadeh}}, \bibinfo {author} {\bibfnamefont {M.~T.}\ \bibnamefont
  {Yamashita}}, \bibinfo {author} {\bibfnamefont {L.}~\bibnamefont {Tomio}},
  \bibinfo {author} {\bibfnamefont {A.}~\bibnamefont {Delfino}}, \ and\
  \bibinfo {author} {\bibfnamefont {T.}~\bibnamefont {Frederico}},\ }\href
  {\doibase 10.1103/PhysRevLett.107.135304} {\bibfield  {journal} {\bibinfo
  {journal} {Phys. Rev. Lett.}\ }\textbf {\bibinfo {volume} {107}},\ \bibinfo
  {pages} {135304} (\bibinfo {year} {2011})},\ \Eprint
  {http://arxiv.org/abs/1101.0378} {arXiv:1101.0378 [physics.atm-clus]}
  \BibitemShut {NoStop}%
\bibitem [{\citenamefont {Frederico}\ \emph {et~al.}(2019)\citenamefont
  {Frederico}, \citenamefont {Paula}, \citenamefont {Delfino}, \citenamefont
  {Yamashita},\ and\ \citenamefont {Tomio}}]{Frederico:2019bnm}%
  \BibitemOpen
  \bibfield  {author} {\bibinfo {author} {\bibfnamefont {T.}~\bibnamefont
  {Frederico}}, \bibinfo {author} {\bibfnamefont {W.}~\bibnamefont {Paula}},
  \bibinfo {author} {\bibfnamefont {A.}~\bibnamefont {Delfino}}, \bibinfo
  {author} {\bibfnamefont {M.~T.}\ \bibnamefont {Yamashita}}, \ and\ \bibinfo
  {author} {\bibfnamefont {L.}~\bibnamefont {Tomio}},\ }\href {\doibase
  10.1007/s00601-019-1514-z} {\bibfield  {journal} {\bibinfo  {journal} {Few
  Body Syst.}\ }\textbf {\bibinfo {volume} {60}},\ \bibinfo {pages} {46}
  (\bibinfo {year} {2019})}\BibitemShut {NoStop}%
\bibitem [{\citenamefont {De~Paula}\ \emph {et~al.}(2020)\citenamefont
  {De~Paula}, \citenamefont {Delfino}, \citenamefont {Frederico},\ and\
  \citenamefont {Tomio}}]{DePaula:2019ryz}%
  \BibitemOpen
  \bibfield  {author} {\bibinfo {author} {\bibfnamefont {W.}~\bibnamefont
  {De~Paula}}, \bibinfo {author} {\bibfnamefont {A.}~\bibnamefont {Delfino}},
  \bibinfo {author} {\bibfnamefont {T.}~\bibnamefont {Frederico}}, \ and\
  \bibinfo {author} {\bibfnamefont {L.}~\bibnamefont {Tomio}},\ }\href
  {\doibase 10.1088/1361-6455/aba9e2} {\bibfield  {journal} {\bibinfo
  {journal} {J. Phys. B}\ }\textbf {\bibinfo {volume} {53}},\ \bibinfo {pages}
  {205301} (\bibinfo {year} {2020})},\ \Eprint
  {http://arxiv.org/abs/1903.10321} {arXiv:1903.10321 [quant-ph]} \BibitemShut
  {NoStop}%
\bibitem [{\citenamefont {Platter}\ \emph {et~al.}(2004)\citenamefont
  {Platter}, \citenamefont {Hammer},\ and\ \citenamefont
  {Meissner}}]{Platter:2004he}%
  \BibitemOpen
  \bibfield  {author} {\bibinfo {author} {\bibfnamefont {L.}~\bibnamefont
  {Platter}}, \bibinfo {author} {\bibfnamefont {H.~W.}\ \bibnamefont {Hammer}},
  \ and\ \bibinfo {author} {\bibfnamefont {U.-G.}\ \bibnamefont {Meissner}},\
  }\href {\doibase 10.1103/PhysRevA.70.052101} {\bibfield  {journal} {\bibinfo
  {journal} {Phys. Rev. A}\ }\textbf {\bibinfo {volume} {70}},\ \bibinfo
  {pages} {052101} (\bibinfo {year} {2004})},\ \Eprint
  {http://arxiv.org/abs/cond-mat/0404313} {arXiv:cond-mat/0404313} \BibitemShut
  {NoStop}%
\bibitem [{\citenamefont {von Stecher}\ \emph {et~al.}(2009)\citenamefont {von
  Stecher}, \citenamefont {D'Incao},\ and\ \citenamefont
  {Greene}}]{StecherNat2009}%
  \BibitemOpen
  \bibfield  {author} {\bibinfo {author} {\bibfnamefont {J.}~\bibnamefont {von
  Stecher}}, \bibinfo {author} {\bibfnamefont {J.~P.}\ \bibnamefont {D'Incao}},
  \ and\ \bibinfo {author} {\bibfnamefont {C.~H.}\ \bibnamefont {Greene}},\
  }\href {\doibase 10.1038/nphys1253} {\bibfield  {journal} {\bibinfo
  {journal} {Nature Physics}\ ,\ \bibinfo {pages} {417}} (\bibinfo {year}
  {2009})}\BibitemShut {NoStop}%
\bibitem [{\citenamefont {Gattobigio}\ \emph {et~al.}(2012)\citenamefont
  {Gattobigio}, \citenamefont {Kievsky},\ and\ \citenamefont
  {Viviani}}]{GattobigioPRA2011}%
  \BibitemOpen
  \bibfield  {author} {\bibinfo {author} {\bibfnamefont {M.}~\bibnamefont
  {Gattobigio}}, \bibinfo {author} {\bibfnamefont {A.}~\bibnamefont {Kievsky}},
  \ and\ \bibinfo {author} {\bibfnamefont {M.}~\bibnamefont {Viviani}},\ }\href
  {\doibase 10.1103/PhysRevA.86.042513} {\bibfield  {journal} {\bibinfo
  {journal} {Phys. Rev. A}\ }\textbf {\bibinfo {volume} {86}},\ \bibinfo
  {pages} {042513} (\bibinfo {year} {2012})}\BibitemShut {NoStop}%
\bibitem [{\citenamefont {Gattobigio}\ \emph {et~al.}(2013)\citenamefont
  {Gattobigio}, \citenamefont {Kievsky},\ and\ \citenamefont
  {Viviani}}]{Gattobigio:2012yky}%
  \BibitemOpen
  \bibfield  {author} {\bibinfo {author} {\bibfnamefont {M.}~\bibnamefont
  {Gattobigio}}, \bibinfo {author} {\bibfnamefont {A.}~\bibnamefont {Kievsky}},
  \ and\ \bibinfo {author} {\bibfnamefont {M.}~\bibnamefont {Viviani}},\ }\href
  {\doibase 10.1007/s00601-013-0630-4} {\bibfield  {journal} {\bibinfo
  {journal} {Few Body Syst.}\ }\textbf {\bibinfo {volume} {54}},\ \bibinfo
  {pages} {1547} (\bibinfo {year} {2013})},\ \Eprint
  {http://arxiv.org/abs/1211.6637} {arXiv:1211.6637 [physics.atm-clus]}
  \BibitemShut {NoStop}%
\bibitem [{\citenamefont {Gattobigio}\ and\ \citenamefont
  {Kievsky}(2014)}]{GattobigioPRA2014}%
  \BibitemOpen
  \bibfield  {author} {\bibinfo {author} {\bibfnamefont {M.}~\bibnamefont
  {Gattobigio}}\ and\ \bibinfo {author} {\bibfnamefont {A.}~\bibnamefont
  {Kievsky}},\ }\href {\doibase 10.1103/PhysRevA.90.012502} {\bibfield
  {journal} {\bibinfo  {journal} {Phys. Rev. A}\ }\textbf {\bibinfo {volume}
  {90}},\ \bibinfo {pages} {012502} (\bibinfo {year} {2014})}\BibitemShut
  {NoStop}%
\bibitem [{\citenamefont {Kievsky}\ \emph
  {et~al.}(2014{\natexlab{a}})\citenamefont {Kievsky}, \citenamefont
  {Gattobigio},\ and\ \citenamefont {Timofeyuk}}]{Kievsky:2014yua}%
  \BibitemOpen
  \bibfield  {author} {\bibinfo {author} {\bibfnamefont {A.}~\bibnamefont
  {Kievsky}}, \bibinfo {author} {\bibfnamefont {M.}~\bibnamefont {Gattobigio}},
  \ and\ \bibinfo {author} {\bibfnamefont {N.~K.}\ \bibnamefont {Timofeyuk}},\
  }\href {\doibase 10.1007/s00601-013-0773-3} {\bibfield  {journal} {\bibinfo
  {journal} {Few Body Syst.}\ }\textbf {\bibinfo {volume} {55}},\ \bibinfo
  {pages} {945} (\bibinfo {year} {2014}{\natexlab{a}})},\ \Eprint
  {http://arxiv.org/abs/1404.7337} {arXiv:1404.7337 [cond-mat.quant-gas]}
  \BibitemShut {NoStop}%
\bibitem [{\citenamefont {Kievsky}\ \emph
  {et~al.}(2014{\natexlab{b}})\citenamefont {Kievsky}, \citenamefont
  {Timofeyuk},\ and\ \citenamefont {Gattobigio}}]{KievskyPRA2014}%
  \BibitemOpen
  \bibfield  {author} {\bibinfo {author} {\bibfnamefont {A.}~\bibnamefont
  {Kievsky}}, \bibinfo {author} {\bibfnamefont {N.~K.}\ \bibnamefont
  {Timofeyuk}}, \ and\ \bibinfo {author} {\bibfnamefont {M.}~\bibnamefont
  {Gattobigio}},\ }\href {\doibase 10.1103/PhysRevA.90.032504} {\bibfield
  {journal} {\bibinfo  {journal} {Phys. Rev. A}\ }\textbf {\bibinfo {volume}
  {90}},\ \bibinfo {pages} {032504} (\bibinfo {year}
  {2014}{\natexlab{b}})}\BibitemShut {NoStop}%
\bibitem [{\citenamefont {\'Alvarez-Rodr\'{\i}guez}\ \emph
  {et~al.}(2016)\citenamefont {\'Alvarez-Rodr\'{\i}guez}, \citenamefont
  {Deltuva}, \citenamefont {Gattobigio},\ and\ \citenamefont
  {Kievsky}}]{RodriguezPRA2016}%
  \BibitemOpen
  \bibfield  {author} {\bibinfo {author} {\bibfnamefont {R.}~\bibnamefont
  {\'Alvarez-Rodr\'{\i}guez}}, \bibinfo {author} {\bibfnamefont
  {A.}~\bibnamefont {Deltuva}}, \bibinfo {author} {\bibfnamefont
  {M.}~\bibnamefont {Gattobigio}}, \ and\ \bibinfo {author} {\bibfnamefont
  {A.}~\bibnamefont {Kievsky}},\ }\href {\doibase 10.1103/PhysRevA.93.062701}
  {\bibfield  {journal} {\bibinfo  {journal} {Phys. Rev. A}\ }\textbf {\bibinfo
  {volume} {93}},\ \bibinfo {pages} {062701} (\bibinfo {year}
  {2016})}\BibitemShut {NoStop}%
\bibitem [{\citenamefont {Deltuva}(2013)}]{Deltuva:2012ig}%
  \BibitemOpen
  \bibfield  {author} {\bibinfo {author} {\bibfnamefont {A.}~\bibnamefont
  {Deltuva}},\ }\href {\doibase 10.1007/s00601-012-0313-6} {\bibfield
  {journal} {\bibinfo  {journal} {Few-Body Systems}\ }\textbf {\bibinfo
  {volume} {54}},\ \bibinfo {pages} {569} (\bibinfo {year} {2013})},\ \Eprint
  {http://arxiv.org/abs/1202.0167} {arXiv:1202.0167 [physics.atom-ph]}
  \BibitemShut {NoStop}%
\bibitem [{\citenamefont {Yamashita}\ \emph {et~al.}(2006)\citenamefont
  {Yamashita}, \citenamefont {Tomio}, \citenamefont {Delfino},\ and\
  \citenamefont {Frederico}}]{Yamashita_2006}%
  \BibitemOpen
  \bibfield  {author} {\bibinfo {author} {\bibfnamefont {M.~T.}\ \bibnamefont
  {Yamashita}}, \bibinfo {author} {\bibfnamefont {L.}~\bibnamefont {Tomio}},
  \bibinfo {author} {\bibfnamefont {A.}~\bibnamefont {Delfino}}, \ and\
  \bibinfo {author} {\bibfnamefont {T.}~\bibnamefont {Frederico}},\ }\href
  {\doibase 10.1209/epl/i2006-10141-6} {\bibfield  {journal} {\bibinfo
  {journal} {Europhysics Letters ({EPL})}\ }\textbf {\bibinfo {volume} {75}},\
  \bibinfo {pages} {555} (\bibinfo {year} {2006})}\BibitemShut {NoStop}%
\bibitem [{\citenamefont {Bazak}\ \emph {et~al.}(2019)\citenamefont {Bazak},
  \citenamefont {Kirscher}, \citenamefont {K\"onig}, \citenamefont
  {Pav\'on~Valderrama}, \citenamefont {Barnea},\ and\ \citenamefont {van
  Kolck}}]{Bazak:2018qnu}%
  \BibitemOpen
  \bibfield  {author} {\bibinfo {author} {\bibfnamefont {B.}~\bibnamefont
  {Bazak}}, \bibinfo {author} {\bibfnamefont {J.}~\bibnamefont {Kirscher}},
  \bibinfo {author} {\bibfnamefont {S.}~\bibnamefont {K\"onig}}, \bibinfo
  {author} {\bibfnamefont {M.}~\bibnamefont {Pav\'on~Valderrama}}, \bibinfo
  {author} {\bibfnamefont {N.}~\bibnamefont {Barnea}}, \ and\ \bibinfo {author}
  {\bibfnamefont {U.}~\bibnamefont {van Kolck}},\ }\href {\doibase
  10.1103/PhysRevLett.122.143001} {\bibfield  {journal} {\bibinfo  {journal}
  {Phys. Rev. Lett.}\ }\textbf {\bibinfo {volume} {122}},\ \bibinfo {pages}
  {143001} (\bibinfo {year} {2019})},\ \Eprint
  {http://arxiv.org/abs/1812.00387} {arXiv:1812.00387 [cond-mat.quant-gas]}
  \BibitemShut {NoStop}%
\bibitem [{\citenamefont {Stipanovi{\'c}}\ \emph {et~al.}(2022)\citenamefont
  {Stipanovi{\'c}}, \citenamefont {Vranje{\v s}~Marki{\'c}},\ and\
  \citenamefont {Boronat}}]{Stipanovic2022}%
  \BibitemOpen
  \bibfield  {author} {\bibinfo {author} {\bibfnamefont {P.}~\bibnamefont
  {Stipanovi{\'c}}}, \bibinfo {author} {\bibfnamefont {L.}~\bibnamefont
  {Vranje{\v s}~Marki{\'c}}}, \ and\ \bibinfo {author} {\bibfnamefont
  {J.}~\bibnamefont {Boronat}},\ }\href@noop {} {\bibfield  {journal} {\bibinfo
   {journal} {Scientific Reports}\ }\textbf {\bibinfo {volume} {12}},\ \bibinfo
  {pages} {10368} (\bibinfo {year} {2022})}\BibitemShut {NoStop}%
\bibitem [{\citenamefont {Horinouchi}\ and\ \citenamefont
  {Ueda}(2015)}]{horinouchi:2015_Phys.Rev.Lett.}%
  \BibitemOpen
  \bibfield  {author} {\bibinfo {author} {\bibfnamefont {Y.}~\bibnamefont
  {Horinouchi}}\ and\ \bibinfo {author} {\bibfnamefont {M.}~\bibnamefont
  {Ueda}},\ }\href {\doibase 10.1103/PhysRevLett.114.025301} {\bibfield
  {journal} {\bibinfo  {journal} {Phys. Rev. Lett.}\ }\textbf {\bibinfo
  {volume} {114}} (\bibinfo {year} {2015}),\
  10.1103/PhysRevLett.114.025301}\BibitemShut {NoStop}%
\bibitem [{\citenamefont {Yamashita}\ \emph {et~al.}(2021)\citenamefont
  {Yamashita}, \citenamefont {Frederico},\ and\ \citenamefont
  {Tomio}}]{Yamashita:2020jvu}%
  \BibitemOpen
  \bibfield  {author} {\bibinfo {author} {\bibfnamefont {M.~T.}\ \bibnamefont
  {Yamashita}}, \bibinfo {author} {\bibfnamefont {T.}~\bibnamefont
  {Frederico}}, \ and\ \bibinfo {author} {\bibfnamefont {L.}~\bibnamefont
  {Tomio}},\ }\href {\doibase 10.1007/s13538-020-00810-6} {\bibfield  {journal}
  {\bibinfo  {journal} {Braz. J. Phys.}\ }\textbf {\bibinfo {volume} {51}},\
  \bibinfo {pages} {277} (\bibinfo {year} {2021})},\ \Eprint
  {http://arxiv.org/abs/2012.03882} {arXiv:2012.03882 [physics.atom-ph]}
  \BibitemShut {NoStop}%
\bibitem [{\citenamefont {{Johansen}}\ \emph {et~al.}(2017)\citenamefont
  {{Johansen}}, \citenamefont {{Desalvo}}, \citenamefont {{Patel}},\ and\
  \citenamefont {{Chin}}}]{ChinNatPhys2017}%
  \BibitemOpen
  \bibfield  {author} {\bibinfo {author} {\bibfnamefont {J.}~\bibnamefont
  {{Johansen}}}, \bibinfo {author} {\bibfnamefont {B.~J.}\ \bibnamefont
  {{Desalvo}}}, \bibinfo {author} {\bibfnamefont {K.}~\bibnamefont {{Patel}}},
  \ and\ \bibinfo {author} {\bibfnamefont {C.}~\bibnamefont {{Chin}}},\ }\href
  {\doibase 10.1038/nphys4130} {\bibfield  {journal} {\bibinfo  {journal}
  {Nature Physics}\ }\textbf {\bibinfo {volume} {13}},\ \bibinfo {pages} {731}
  (\bibinfo {year} {2017})},\ \Eprint {http://arxiv.org/abs/1612.05169}
  {arXiv:1612.05169 [cond-mat.quant-gas]} \BibitemShut {NoStop}%
\bibitem [{\citenamefont {Chapurin}\ \emph {et~al.}(2019)\citenamefont
  {Chapurin}, \citenamefont {Xie}, \citenamefont {Van~de Graaff}, \citenamefont
  {Popowski}, \citenamefont {D'Incao}, \citenamefont {Julienne}, \citenamefont
  {Ye},\ and\ \citenamefont {Cornell}}]{Chapurin_2019}%
  \BibitemOpen
  \bibfield  {author} {\bibinfo {author} {\bibfnamefont {R.}~\bibnamefont
  {Chapurin}}, \bibinfo {author} {\bibfnamefont {X.}~\bibnamefont {Xie}},
  \bibinfo {author} {\bibfnamefont {M.~J.}\ \bibnamefont {Van~de Graaff}},
  \bibinfo {author} {\bibfnamefont {J.~S.}\ \bibnamefont {Popowski}}, \bibinfo
  {author} {\bibfnamefont {J.~P.}\ \bibnamefont {D'Incao}}, \bibinfo {author}
  {\bibfnamefont {P.~S.}\ \bibnamefont {Julienne}}, \bibinfo {author}
  {\bibfnamefont {J.}~\bibnamefont {Ye}}, \ and\ \bibinfo {author}
  {\bibfnamefont {E.~A.}\ \bibnamefont {Cornell}},\ }\href {\doibase
  10.1103/PhysRevLett.123.233402} {\bibfield  {journal} {\bibinfo  {journal}
  {Phys. Rev. Lett.}\ }\textbf {\bibinfo {volume} {123}},\ \bibinfo {pages}
  {233402} (\bibinfo {year} {2019})}\BibitemShut {NoStop}%
\bibitem [{\citenamefont {Xie}\ \emph {et~al.}(2020)\citenamefont {Xie},
  \citenamefont {Van~de Graaff}, \citenamefont {Chapurin}, \citenamefont
  {Frye}, \citenamefont {Hutson}, \citenamefont {D'Incao}, \citenamefont
  {Julienne}, \citenamefont {Ye},\ and\ \citenamefont
  {Cornell}}]{CornellPRL2020}%
  \BibitemOpen
  \bibfield  {author} {\bibinfo {author} {\bibfnamefont {X.}~\bibnamefont
  {Xie}}, \bibinfo {author} {\bibfnamefont {M.~J.}\ \bibnamefont {Van~de
  Graaff}}, \bibinfo {author} {\bibfnamefont {R.}~\bibnamefont {Chapurin}},
  \bibinfo {author} {\bibfnamefont {M.~D.}\ \bibnamefont {Frye}}, \bibinfo
  {author} {\bibfnamefont {J.~M.}\ \bibnamefont {Hutson}}, \bibinfo {author}
  {\bibfnamefont {J.~P.}\ \bibnamefont {D'Incao}}, \bibinfo {author}
  {\bibfnamefont {P.~S.}\ \bibnamefont {Julienne}}, \bibinfo {author}
  {\bibfnamefont {J.}~\bibnamefont {Ye}}, \ and\ \bibinfo {author}
  {\bibfnamefont {E.~A.}\ \bibnamefont {Cornell}},\ }\href {\doibase
  10.1103/PhysRevLett.125.243401} {\bibfield  {journal} {\bibinfo  {journal}
  {Phys. Rev. Lett.}\ }\textbf {\bibinfo {volume} {125}},\ \bibinfo {pages}
  {243401} (\bibinfo {year} {2020})}\BibitemShut {NoStop}%
\bibitem [{\citenamefont {Naidon}\ and\ \citenamefont
  {Endo}(2017)}]{Naidon:2016dpf}%
  \BibitemOpen
  \bibfield  {author} {\bibinfo {author} {\bibfnamefont {P.}~\bibnamefont
  {Naidon}}\ and\ \bibinfo {author} {\bibfnamefont {S.}~\bibnamefont {Endo}},\
  }\href {\doibase 10.1088/1361-6633/aa50e8} {\bibfield  {journal} {\bibinfo
  {journal} {Rept. Prog. Phys.}\ }\textbf {\bibinfo {volume} {80}},\ \bibinfo
  {pages} {056001} (\bibinfo {year} {2017})},\ \Eprint
  {http://arxiv.org/abs/1610.09805} {arXiv:1610.09805 [quant-ph]} \BibitemShut
  {NoStop}%
\bibitem [{\citenamefont {Greene}\ \emph {et~al.}(2017)\citenamefont {Greene},
  \citenamefont {Giannakeas},\ and\ \citenamefont
  {Perez-Rios}}]{Greene:2017cik}%
  \BibitemOpen
  \bibfield  {author} {\bibinfo {author} {\bibfnamefont {C.~H.}\ \bibnamefont
  {Greene}}, \bibinfo {author} {\bibfnamefont {P.}~\bibnamefont {Giannakeas}},
  \ and\ \bibinfo {author} {\bibfnamefont {J.}~\bibnamefont {Perez-Rios}},\
  }\href {\doibase 10.1103/RevModPhys.89.035006} {\bibfield  {journal}
  {\bibinfo  {journal} {Rev. Mod. Phys.}\ }\textbf {\bibinfo {volume} {89}},\
  \bibinfo {pages} {035006} (\bibinfo {year} {2017})},\ \Eprint
  {http://arxiv.org/abs/1704.02029} {arXiv:1704.02029 [cond-mat.quant-gas]}
  \BibitemShut {NoStop}%
\bibitem [{\citenamefont {Varga}\ and\ \citenamefont
  {Suzuki}(1995)}]{varga:1995_Phys.Rev.C}%
  \BibitemOpen
  \bibfield  {author} {\bibinfo {author} {\bibfnamefont {K.}~\bibnamefont
  {Varga}}\ and\ \bibinfo {author} {\bibfnamefont {Y.}~\bibnamefont {Suzuki}},\
  }\href {\doibase 10.1103/PhysRevC.52.2885} {\bibfield  {journal} {\bibinfo
  {journal} {Phys. Rev. C}\ }\textbf {\bibinfo {volume} {52}},\ \bibinfo
  {pages} {2885} (\bibinfo {year} {1995})}\BibitemShut {NoStop}%
\bibitem [{\citenamefont {Suzuki}\ and\ \citenamefont
  {Varga}(1998)}]{suzuki:1998_}%
  \BibitemOpen
  \bibfield  {author} {\bibinfo {author} {\bibfnamefont {Y.}~\bibnamefont
  {Suzuki}}\ and\ \bibinfo {author} {\bibfnamefont {K.}~\bibnamefont {Varga}},\
  }\href@noop {} {\emph {\bibinfo {title} {Stochastic Variational Approach to
  Quantum-Mechanical Few-Body Problems}}}\ (\bibinfo  {publisher}
  {{Springer}},\ \bibinfo {address} {{Berlin; New York}},\ \bibinfo {year}
  {1998})\BibitemShut {NoStop}%
\bibitem [{\citenamefont {Deltuva}\ \emph {et~al.}(2020)\citenamefont
  {Deltuva}, \citenamefont {Gattobigio}, \citenamefont {Kievsky},\ and\
  \citenamefont {Viviani}}]{deltuva:2020_Phys.Rev.C}%
  \BibitemOpen
  \bibfield  {author} {\bibinfo {author} {\bibfnamefont {A.}~\bibnamefont
  {Deltuva}}, \bibinfo {author} {\bibfnamefont {M.}~\bibnamefont {Gattobigio}},
  \bibinfo {author} {\bibfnamefont {A.}~\bibnamefont {Kievsky}}, \ and\
  \bibinfo {author} {\bibfnamefont {M.}~\bibnamefont {Viviani}},\ }\href
  {\doibase 10.1103/PhysRevC.102.064001} {\bibfield  {journal} {\bibinfo
  {journal} {Phys. Rev. C}\ }\textbf {\bibinfo {volume} {102}},\ \bibinfo
  {pages} {064001} (\bibinfo {year} {2020})}\BibitemShut {NoStop}%
\bibitem [{\citenamefont {Kievsky}\ \emph {et~al.}(2021)\citenamefont
  {Kievsky}, \citenamefont {Gattobigio}, \citenamefont {Girlanda},\ and\
  \citenamefont {Viviani}}]{kievsky:2021_Annu.Rev.Nucl.Part.Sci.}%
  \BibitemOpen
  \bibfield  {author} {\bibinfo {author} {\bibfnamefont {A.}~\bibnamefont
  {Kievsky}}, \bibinfo {author} {\bibfnamefont {M.}~\bibnamefont {Gattobigio}},
  \bibinfo {author} {\bibfnamefont {L.}~\bibnamefont {Girlanda}}, \ and\
  \bibinfo {author} {\bibfnamefont {M.}~\bibnamefont {Viviani}},\ }\href
  {\doibase 10.1146/annurev-nucl-102419-032845} {\bibfield  {journal} {\bibinfo
   {journal} {Annu. Rev. Nucl. Part. Sci.}\ }\textbf {\bibinfo {volume} {71}},\
  \bibinfo {pages} {465} (\bibinfo {year} {2021})}\BibitemShut {NoStop}%
\bibitem [{\citenamefont {Combescure}\ \emph {et~al.}(2007)\citenamefont
  {Combescure}, \citenamefont {Khare}, \citenamefont {Raina}, \citenamefont
  {Richard},\ and\ \citenamefont {Weydert}}]{combescure:2007_Int.J.Mod.Phys.B}%
  \BibitemOpen
  \bibfield  {author} {\bibinfo {author} {\bibfnamefont {M.}~\bibnamefont
  {Combescure}}, \bibinfo {author} {\bibfnamefont {A.}~\bibnamefont {Khare}},
  \bibinfo {author} {\bibfnamefont {A.~K.}\ \bibnamefont {Raina}}, \bibinfo
  {author} {\bibfnamefont {J.-M.}\ \bibnamefont {Richard}}, \ and\ \bibinfo
  {author} {\bibfnamefont {C.}~\bibnamefont {Weydert}},\ }\href {\doibase
  10.1142/S0217979207037855} {\bibfield  {journal} {\bibinfo  {journal}
  {International Journal of Modern Physics B}\ }\textbf {\bibinfo {volume}
  {21}},\ \bibinfo {pages} {3765} (\bibinfo {year} {2007})}\BibitemShut
  {NoStop}%
\bibitem [{\citenamefont {{Secker}}\ \emph {et~al.}(2021)\citenamefont
  {{Secker}}, \citenamefont {{Ahmed-Braun}}, \citenamefont {{Mestrom}},\ and\
  \citenamefont {{Kokkelmans}}}]{KokkelmansPRA2021}%
  \BibitemOpen
  \bibfield  {author} {\bibinfo {author} {\bibfnamefont {T.}~\bibnamefont
  {{Secker}}}, \bibinfo {author} {\bibfnamefont {D.~J.~M.}\ \bibnamefont
  {{Ahmed-Braun}}}, \bibinfo {author} {\bibfnamefont {P.~M.~A.}\ \bibnamefont
  {{Mestrom}}}, \ and\ \bibinfo {author} {\bibfnamefont {S.~J.~J.~M.~F.}\
  \bibnamefont {{Kokkelmans}}},\ }\href {\doibase 10.1103/PhysRevA.103.052805}
  {\bibfield  {journal} {\bibinfo  {journal} {Physical Review A}\ }\textbf
  {\bibinfo {volume} {103}},\ \bibinfo {eid} {052805} (\bibinfo {year}
  {2021})},\ \Eprint {http://arxiv.org/abs/2011.14332} {arXiv:2011.14332
  [cond-mat.quant-gas]} \BibitemShut {NoStop}%
\end{thebibliography}
\end{document}